\documentclass[eqsecnum,preprint,prd,aps,nofootinbib]{revtex4}
\usepackage{amsmath}
\usepackage{graphics}
\newcommand{\be}{\begin{equation}}
\newcommand{\ee}{\end{equation}}
\newcommand{\ba}{\begin{eqnarray}}
\newcommand{\ea}{\end{eqnarray}}
\newcommand{\bc}{\begin{center}}
\newcommand{\ec}{\end{center}}
\def\lvec#1{\vbox{\ialign{##\crcr$\leftarrow$\crcr\noalign{
 \kern-1pt\nointerlineskip}$\hfil\displaystyle{#1}\hfil$\crcr}}}
\def\rvec#1{\vbox{\ialign{##\crcr$\rightarrow$\crcr\noalign{
 \kern-1pt\nointerlineskip}$\hfil\displaystyle{#1}\hfil$\crcr}}}

\begin{document}
\begin{center}
\bibliographystyle{article}

{\Large \textsc{From Peierls brackets to a generalized 
Moyal bracket for type-I gauge theories}}

\end{center}
\vspace{0.4cm}


\date{\today}

\author{Giampiero Esposito,$^{1,2}$ \thanks{%
Electronic address: giampiero.esposito@na.infn.it} 
Cosimo Stornaiolo$^{1,2}$ \thanks{%
Electronic address: cosmo@na.infn.it}}
\affiliation{${\ }^{1}$Istituto Nazionale di Fisica Nucleare, 
Sezione di Napoli,\\
Complesso Universitario di Monte S. Angelo, Via Cintia, Edificio N', 80126
Napoli, Italy\\
${\ }^{2}$Dipartimento di Scienze Fisiche, Complesso Universitario di Monte
S. Angelo,\\
Via Cintia, Edificio N', 80126 Napoli, Italy}

\begin{abstract}
In the space-of-histories approach to gauge fields and their quantization, the
Maxwell, Yang--Mills and gravitational field are well known to share the
property of being type-I theories, i.e. Lie brackets of the vector fields
which leave the action functional invariant are linear combinations of such
vector fields, with coefficients of linear combination given by structure
constants. The corresponding gauge-field operator in the functional integral 
for the in-out amplitude is an invertible second-order differential operator.
For such an operator, we consider advanced and retarded Green functions 
giving rise to a Peierls bracket among group-invariant functionals. Our 
Peierls bracket is a Poisson bracket on the space of all group-invariant
functionals in two cases only: either the gauge-fixing is arbitrary but
the gauge fields lie on the dynamical sub-space; or the gauge-fixing is a
linear functional of gauge fields, which are generic points of the space
of histories. In both cases, the resulting Peierls bracket is proved to be
gauge-invariant by exploiting the manifestly covariant formalism.
Moreover, on quantization, a gauge-invariant Moyal bracket is defined that 
reduces to $i {\hbar}$ times the Peierls bracket to lowest order in 
${\hbar}$.
\end{abstract}
\maketitle
\bigskip
\vspace{2cm}

\section{Introduction}
The modern formulations of quantum field theory and quantum gravity still
reflect two basic attitudes: either one follows a Lagrangian path, starting 
from an action functional with the associated functional integral formulation
\cite{DeWi03}, or the Hamiltonian road to quantization is chosen, with the
associated constraint analysis \cite{Dira01} and functional differential
equations (the latter being extremely difficult, especially for gravitation).
The two approaches are not obviously equivalent in all cases \cite{Hall91},
and one can indeed build a sum-over-histories which does not solve the 
constraint equations of the quantum theory via Hamiltonian methods
\cite{DeWi98}. A further relevant example is provided by quantum
supergravity: the supersymmetry constraints lead to equations solved 
exactly by a wave function which is finite \cite{Deat96}, whereas the
analysis of counterterms within the framework of covariant perturbation
methods shows no hope for finiteness of quantum supergravity
\cite{Dese00, Stel02}.

It is therefore important to re-assess the foundations of covariant methods
on the one hand, and their relation with `covariant' formulations of
Hamiltonian quantization on the other hand \cite{Ozak05}. 
In particular, a cornerstone
of the space-time approach to quantum field theory \cite{DeWi84} is the 
Peierls bracket \cite{Peie52}, that makes it possible to have a Poisson
bracket which is completely invariant under the (proper) gauge group
\cite{DeWi03} (this being an infinite-dimensional Lie group). For our 
purposes, the framework we are interested in can be described as follows
\cite{DeWi03, DeWi65}.

To begin, an action functional $S$ is given, which is a real-valued 
functional defined on the space $\Phi$ of field histories. On $\Phi$,
a set of vector fields $Q_{\alpha}$ exist which leave the action invariant,
i.e.
\begin{equation}
Q_{\alpha}S=0,
\label{(1.1)}
\end{equation}
and having components $Q_{\; \alpha}^{i}$. Lower case Latin indices are 
used for components of fields $\varphi^{i}$ (e.g. $\varphi^{i}=A_{\mu}(x)$,
or $A_{\mu}^{\alpha}(x)$, or $g_{\mu \nu}(x)$), while Greek indices from the
beginning of the alphabet are Lie-algebra indices. Whenever Latin or Greek
indices are summed over, this means contraction jointly with integration,
e.g. 
$$
S_{,ij'}u^{j'}=\int {\delta^{2}S \over \delta \varphi^{i}(x)
\delta \varphi^{j}(x')}u^{j}(x')dx',
$$
while infinitesimal gauge transformations read
\begin{equation}
\delta \varphi^{i}=\int Q_{\; \alpha}^{i}(x,x')
\delta \xi^{\alpha}(x')dx'=Q_{\; \alpha}^{i}\delta \xi^{\alpha}.
\label{(1.2)}
\end{equation}
The vector fields $Q_{\alpha}$ are linearly independent and have Lie 
brackets satisfying
\begin{equation}
\Bigr[Q_{\alpha},Q_{\beta}\Bigr]=C_{\; \alpha \beta}^{\gamma}
Q_{\gamma}+S_{,i}T_{\; \; \; \alpha \beta}^{i \cdot}.
\label{(1.3)}
\end{equation}
For type-I theories, that we consider hereafter and include Maxwell,
Yang--Mills and general relativity, the $C_{\; \; \alpha \beta}^{\gamma}$
are structure constants in that
\begin{equation}
{\delta \over \delta \varphi^{i}}C_{\; \alpha \beta}^{\gamma}
=C_{\; \alpha \beta,i}^{\gamma}=0,
\label{(1.4)}
\end{equation}
and $T_{\; \; \; \alpha \beta}^{i \cdot}$ vanishes as well. The
components $Q_{\; \alpha}^{i}$ can be taken to depend linearly on the
fields $\varphi^{i}$, i.e.
\begin{equation}
Q_{\; \alpha,jk}^{i}=0,
\label{(1.5)}
\end{equation}
since, on acting with $Q_{\rho}$ on both sides of the Lie bracket
\begin{equation}
\Bigr[Q_{\alpha},Q_{\beta}\Bigr]=C_{\; \alpha \beta}^{\gamma}
Q_{\gamma},
\label{(1.6)}
\end{equation}
the resulting second functional derivatives $Q_{\; \alpha,jk}^{i}$
multiply vanishing terms like $Q_{\; \beta}^{j}Q_{\; \gamma}^{k}$
weighted with opposite signs.
Moreover, the $Q_{\; \alpha}^{i}$ are a sum of Dirac delta $\delta(x,x')$ 
and/or their first derivatives, multiplied by local functions of 
$\varphi^{i}$ and their first derivatives. For example, in the infinitesimal
gauge transformation for Maxwell theory: 
\begin{equation}
\delta A_{\mu}(x)=\partial_{\mu}\delta \xi(x)
=\int -\delta_{,\mu}(x,x')\delta \xi(x')dx',
\label{(1.7)}
\end{equation}
$Q_{\; \alpha}^{i}$ reduces to $Q_{\mu}(x,x')=-\delta_{,\mu}(x,x')$.

By virtue of Eq. (1.6) the proper gauge group 
$\cal G$ obtained by exponentiating 
the transformations (1.2) decomposes $\Phi$ into sub-spaces, called
orbits, to which the vector fields $Q_{\alpha}$ are 
tangent \cite{DeWi03}. The space
$\Phi$ is a principal fibre bundle over ${{\Phi}/{\cal G}}$, and the 
base space ${{\Phi}/{\cal G}}$ is the space of orbits. Fibre-adapted
coordinates consist of abstract coordinates $I^{A}$ which label fibres
of $\Phi \rightarrow {{\Phi}/{\cal G}}$, jointly with $P^{\alpha}$
coordinates which label points within each fibre. The $P^{\alpha}$
correspond to a choice of gauge-fixing functional $P^{\alpha}$ in the
functional integral for the $\langle {\rm out} | {\rm in} \rangle$ 
amplitude. On going from $(I^{A},P^{\alpha})$ coordinates to field
variables $\varphi^{i}$, the loop expansion of the 
$\langle {\rm out}| {\rm in} \rangle$ amplitude involves eventually
two invertible operators, i.e. the gauge-field operator 
\begin{equation}
F_{ij}=S_{,ij}+P_{\; ,i}^{\alpha} \; \omega_{\alpha \beta} \;
P_{\; ,j}^{\beta},
\label{(1.8)}
\end{equation}
$\omega_{\alpha \beta}$ being taken to be, for the sake of manifest
covariance, a $\varphi$-dependent local distribution obeying the
gauge transformation law (see, however, 
the end of section 3 for an alternative scheme)
\begin{equation}
\delta \omega_{\alpha \beta}=\omega_{\alpha \beta ,i}
Q_{\; \gamma}^{i} \delta \xi^{\gamma}
=-\Bigr(\omega_{\delta \beta}C_{\; \gamma \alpha}^{\delta}
+\omega_{\alpha \delta}C_{\; \beta \gamma}^{\delta}\Bigr)
\delta \xi^{\gamma},
\label{(1.9)}
\end{equation}
as well as the ghost operator
\begin{equation}
{\widehat F}_{\; \beta}^{\alpha}=Q_{\beta}P^{\alpha}\
=P_{\; ,i}^{\alpha} \; Q_{\; \beta}^{i}.
\label{(1.10)}
\end{equation}
Since $P^{\alpha}$ is chosen in such a way that $F_{ij}$ and
${\widehat F}_{\; \beta}^{\alpha}$ are invertible, one can consider their
Green functions $G^{ij}$ and $G^{\alpha \beta}$, for which
\begin{equation}
F_{ij}G^{jk}=-\delta_{i}^{\; k},
\label{(1.11)}
\end{equation}
\begin{equation}
{\widehat F}_{\alpha \beta}G^{\beta \gamma}
=-\delta_{\alpha}^{\; \gamma}.
\label{(1.12)}
\end{equation}
From the advanced and retarded Green functions of $F_{ij}$, hereafter
denoted as $G^{+ij}$ and $G^{-ij}$, respectively, one can build the
super-commutator function
\begin{equation}
{\widetilde G}^{ij} \equiv G^{+ij}-G^{-ij},
\label{(1.13)}
\end{equation}
and hence the Peierls bracket
\begin{equation}
(A,B) \equiv A_{,i}{\widetilde G}^{ij}B_{,j}
=\int d^{4}x \int d^{4}y {\delta A \over \delta \varphi^{i}(x)}
{\widetilde G}^{ij}(x,y){\delta B \over \delta \varphi^{j}(y)},
\label{(1.14)}
\end{equation}
where $A$ and $B$ are any pair of gauge-invariant functionals of the
fields, i.e.
\begin{equation}
Q_{\alpha}A=0 \Longrightarrow A_{,i}Q_{\; \alpha}^{i}=0, \;
Q_{\alpha}B=0 \Longrightarrow B_{,i}Q_{\; \alpha}^{i}=0.
\label{(1.15)}
\end{equation}
Since the gauge-field operator $F_{ij}$ is the naturally occurring
invertible operator in the quantum theory of gauge fields from the point
of view of functional-integral approach, we have been led to define the
Peierls bracket as in Eq. (1.14). The same definition has been proposed
in Ref. \cite{Maro94}.

Section 2 proves under which conditions Eq. (1.14) defines indeed a
Poisson bracket on the space of all gauge-invariant functionals obeying
Eq. (1.15). Section 3 proves gauge invariance of the Peierls bracket 
(1.14). A covariant Moyal bracket on the space of histories is proposed
in section 4, while concluding remarks and open problems are presented
in section 5. 

\section{Jacobi identity for the Peierls bracket}

Since advanced and retarded Green functions for type-I theories are
related by
\begin{equation}
G^{+ij}=G^{-ji},
\label{(2.1)}
\end{equation}
which implies ${\widetilde G}^{ij}=-{\widetilde G}^{ji}$, the antisymmetry
of (1.14), i.e. 
\begin{equation}
(A,B)=-(B,A),
\label{(2.2)}
\end{equation}
follows immediately from the definition. Bilinearity is also obtained 
at once from (1.14):
\begin{equation}
(A,B+C)=A_{,i}{\widetilde G}^{ij}(B_{,j}+C_{,j})=(A,B)+(A,C).
\label{(2.3)}
\end{equation}
The only non-trivial task is the verification of the Jacobi identity.
Indeed, one finds \cite{Bimo03}
\begin{eqnarray}
P(A,B,C) & \equiv & (A,(B,C))+(B,(C,A))+(C,(A,B))
=A_{,il}B_{,j}C_{,k}\Bigr({\widetilde G}^{ij}{\widetilde G}^{kl}
+{\widetilde G}^{jl}{\widetilde G}^{ki}\Bigr) \nonumber \\
&+& A_{,i}B_{,jl}C_{,k}\Bigr({\widetilde G}^{jk}{\widetilde G}^{il}
+{\widetilde G}^{kl}{\widetilde G}^{ij}\Bigr)
+A_{,i}B_{,j}C_{,kl}\Bigr({\widetilde G}^{ki}
{\widetilde G}^{jl}+{\widetilde G}^{il}{\widetilde G}^{jk}\Bigr)
\nonumber \\
&+& A_{,i}B_{,j}C_{,k}\Bigr({\widetilde G}^{il}
{\widetilde G}_{\; \; \; ,l}^{jk}
+{\widetilde G}^{jl}{\widetilde G}_{\; \; \; ,l}^{ki}
+{\widetilde G}^{kl}{\widetilde G}_{\; \; \; ,l}^{ij}\Bigr).
\label{(2.4)}
\end{eqnarray}
The antisymmetry of the supercommutator function ${\widetilde G}^{ij}$,
jointly with commutation of functional derivatives: $T_{,il}=T_{,li}$ for
all $T=A,B,C$, implies that the first three terms on the last equality
in (2.4) vanish. For example, one finds
\begin{eqnarray}
\; & \; & A_{,il}B_{,j}C_{,k}\Bigr({\widetilde G}^{ij}
{\widetilde G}^{kl}+{\widetilde G}^{jl}{\widetilde G}^{ki}\Bigr)
=A_{,li}B_{,j}C_{,k}\Bigr({\widetilde G}^{lj}{\widetilde G}^{ki}
+{\widetilde G}^{ji}{\widetilde G}^{kl}\Bigr) \nonumber \\
&=& - A_{,il}B_{,j}C_{,k}\Bigr({\widetilde G}^{jl}
{\widetilde G}^{ki}+{\widetilde G}^{ij}{\widetilde G}^{kl}\Bigr)=0,
\label{(2.5)}
\end{eqnarray}
and an entirely analogous procedure can be applied to the terms 
containing the second functional derivatives $B_{,jl}$ and $C_{,kl}$.

The last term in (2.4) requires more labour because it contains 
functional derivatives of ${\widetilde G}^{ij}$. To begin note that,
from infinitesimal variations of Eq. (1.11), one finds
\begin{equation}
\delta G^{\pm}=G^{\pm}(\delta F)G^{\pm},
\label{(2.6)}
\end{equation}
and hence, for any Green function of $F_{ij}$,
\begin{eqnarray}
\; & \; & G_{\; \; \; ,k}^{lm}=G^{li}F_{ij,k}G^{jm} 
= G^{li}S_{,ijk}G^{jm}+G^{li}P_{\; ,i}^{\alpha}
\omega_{\alpha \beta,k}P_{\; ,j}^{\beta}G^{jm} \nonumber \\
&+& G^{li}P_{\; ,ik}^{\alpha}\omega_{\alpha \beta}P_{\; ,j}^{\beta}G^{jm}
+G^{li}P_{\; ,i}^{\alpha}\omega_{\alpha \beta}P_{\; ,jk}^{\beta}G^{jm}.
\label{(2.7)}
\end{eqnarray}
In Eq. (2.7), the terms involving third functional derivatives of the 
action give vanishing contribution $U(A,B,C)$ to the Jacobi identity
(2.4), because \cite{Bimo03}
\begin{eqnarray}
\; & \; & U(A,B,C)=A_{,i}B_{,j}C_{,k}\Bigr[(G^{+ia}-G^{-ia})
(G^{+jb}G^{-kc}-G^{-jb}G^{+kc})\nonumber \\
&+& (G^{+jb}-G^{-jb})(G^{+kc}G^{-ia}-G^{-kc}G^{+ia}) \nonumber \\
&+& (G^{+kc}-G^{-kc})(G^{+ia}G^{-jb}-G^{-ia}G^{+jb})\Bigr]S_{,abc}.
\label{(2.8)}
\end{eqnarray}
This sum vanishes since it involves six pairs of triple products of Green
functions with opposite signs, i.e.
$$
G^{+ia}G^{+jb}G^{-kc}, \; G^{-ia}G^{-jb}G^{+kc}, \; 
G^{+jb}G^{+kc}G^{-ia}, \; G^{-jb}G^{-kc}G^{+ia}, 
$$
$$
G^{+kc}G^{+ia}G^{-jb}, \; G^{-kc}G^{-ia}G^{+jb}.
$$

In the evaluation of the Jacobi identity we therefore deal eventually,
from Eqs. (2.4) and (2.7), with three terms like
\begin{eqnarray}
\; & \; & V(A,B,C)=A_{,i}B_{,j}C_{,k}{\widetilde G}^{kl}
\Bigr[G^{+ir}P_{\; ,r}^{\alpha} \omega_{\alpha \beta,l}
P_{\; ,s}^{\beta}G^{+sj} \nonumber \\
&+& G^{+ir}P_{\; ,rl}^{\alpha}\omega_{\alpha \beta}
P_{\; ,s}^{\beta}G^{+sj}+G^{+ir}P_{\; ,r}^{\alpha}
\omega_{\alpha \beta}P_{\; ,sl}^{\beta}G^{+sj} \nonumber \\
&+& G^{-ir}P_{\; ,r}^{\alpha} \omega_{\alpha \beta,l}
P_{\; ,s}^{\beta}G^{-sj} \nonumber \\
&+& G^{-ir}P_{\; ,rl}^{\alpha} \omega_{\alpha \beta}
P_{\; ,s}^{\beta}G^{-sj}
+G^{-ir}P_{\; ,r}^{\alpha} \omega_{\alpha \beta}
P_{\; ,sl}^{\beta}G^{-sj}\Bigr].
\label{(2.9)}
\end{eqnarray}
If $\omega_{\alpha \beta}$ were taken to be just a non-singular, symmetric,
continuous matrix, independent of field variables, $V(A,B,C)$ would vanish,
and hence the Jacobi identity would be fulfilled, provided that either
\begin{equation}
Z_{,i}G^{ir}P_{\; ,r}^{\alpha}=0, \; {\rm with} \; 
Z=A,B,C,
\label{(2.10)}
\end{equation}
or 
\begin{equation}
P_{\; ,ij}^{\alpha}=0,
\label{(2.11)}
\end{equation}
where (2.10) and (2.11) are sufficient conditions for Jacobi to hold. If
instead $\omega_{\alpha \beta}$ is taken to be a $\varphi$-dependent
local distribution, as in Sec. 1 following Ref. \cite{DeWi81}, the desired
sufficient condition is expressed by Eq. (2.10) only, so that, in both
cases, we are led to consider functional identities involving the left-hand
side of Eq. (2.10). For this purpose, we first note that
\begin{equation}
F_{ik}Q_{\; \alpha}^{k}=S_{,ik}Q_{\; \alpha}^{k}
+P_{\; ,i}^{\beta}\omega_{\beta \gamma}P_{\; ,k}^{\gamma}
Q_{\; \alpha}^{k} 
= -S_{,k}Q_{\; \alpha,i}^{k}
+P_{\; ,i}^{\beta} \omega_{\beta \gamma} 
{\widehat F}_{\; \alpha}^{\gamma},
\label{(2.12)}
\end{equation}
from the gauge invariance of the action (Eq. (1.1)) and from the definition
of ghost operator (Eq. (1.10)). By acting on both sides of Eq. (2.12) with
any Green function of the gauge-field operator one finds, from Eq. (1.11),
\begin{equation}
Q_{\; \alpha}^{i}=G^{ij}S_{,k}Q_{\; \alpha ,j}^{k}
-G^{ir}P_{\; ,r}^{\rho}\omega_{\rho \gamma}
{\widehat F}_{\; \alpha}^{\gamma}.
\label{(2.13)}
\end{equation}
At this stage, we can exploit Eq. (1.15) for any gauge-invariant functional
$Z[\varphi]$ to find, from the identity (2.13),
\begin{equation}
0=Z_{,i}Q_{\; \alpha}^{i}=Z_{,i}G^{ij}S_{,k}Q_{\; \alpha,j}^{k}
-Z_{,i}G^{ir}P_{\; ,r}^{\rho}\omega_{\rho \gamma}
{\widehat F}_{\; \alpha}^{\gamma},
\label{(2.14)}
\end{equation}
and hence, from Eq. (1.12),
\begin{equation}
0=Z_{,i}Q_{\; \alpha}^{i}G^{\alpha \beta}
=Z_{,i}G^{ij}S_{,k}Q_{\; \alpha,j}^{k}G^{\alpha \beta}
+Z_{,i}G^{ir}P_{\; ,r}^{\rho}\omega_{\rho}^{\; \beta}.
\label{(2.15)}
\end{equation}
Eventually, the definition of inverse matrix 
\begin{equation}
\omega_{\rho \beta}\omega^{\beta \alpha}
=\omega_{\rho}^{\; \beta} \; \omega_{\beta}^{\; \alpha}
=\delta_{\rho}^{\; \alpha}
\label{(2.16)}
\end{equation}
yields, from Eq. (2.15), the desired identity in the form 
\begin{equation}
Z_{,i}G^{ir}P_{\; ,r}^{\alpha}=-Z_{,i}G^{ij}S_{,k}
Q_{\; \gamma,j}^{k} G^{\gamma \beta} \omega_{\beta}^{\; \alpha}.
\label{(2.17)}
\end{equation}
Thus, the sufficient condition (2.10) holds if and only if the gauge 
fields lie on the dynamical subspace where the action is stationary, i.e.
\begin{equation}
S_{,k}=0.
\label{(2.18)}
\end{equation}
This is a very restrictive condition for us to be able to prove the
Jacobi identity with a field-dependent matrix $\omega_{\alpha \beta}$
in the gauge-field operator (1.8). If we relax this assumption and
just work with a non-singular, symmetric $\omega_{\alpha \beta}$,
we obtain instead, from Eq. (2.9), the sufficient condition (2.11).
In other words, {\it linear covariant gauges are naturally picked out if
Eq. (1.14) is required to define a good Peierls bracket which obeys
the Jacobi identity}.

Last, the Leibniz rule
\begin{equation}
(A,BC)=(A,B)C+B(A,C)
\label{(2.19)}
\end{equation}
is immediately obtained from (1.14) and from the Leibniz rule for
functional derivatives in type-I theories, i.e.
\begin{equation}
(BC)_{,j}=B_{,j}C+BC_{,j}.
\label{(2.20)}
\end{equation}

\section{Gauge invariance of the Peierls bracket}

When the gauge fields are subject to infinitesimal gauge transformations
according to Eq. (1.2), the Peierls bracket (1.14) follows the gauge
transformation law
\begin{equation}
\delta(A,B)=(\delta A_{,i}){\widetilde G}^{ij}B_{,j}
+A_{,i}(\delta {\widetilde G}^{ij})B_{,j}
+A_{,i}{\widetilde G}^{ij}(\delta B_{,j}),
\label{(3.1)}
\end{equation}
where 
\begin{equation}
\delta A_{,i}=A_{,ik}\delta \varphi^{k}=A_{,ik}Q_{\; \alpha}^{k}
\delta \xi^{\alpha}=-A_{,k}Q_{\; \alpha ,i}^{k} \delta \xi^{\alpha}
\label{(3.2)}
\end{equation}
from the gauge-invariance condition in Eq. (1.15), and the same holds
for $\delta B_{,j}$. As far as the gauge-transformation law of the
supercommutator function ${\widetilde G}^{ij}$ is concerned, this is
obtained by imposing Eq. (1.9) jointly with \cite{DeWi81}
\begin{equation}
\delta S_{,ij}=-\Bigr(S_{,kj}Q_{\; \alpha,i}^{k}
+S_{,ik}Q_{\; \alpha,j}^{k}\Bigr)\delta \xi^{\alpha},
\label{(3.3)}
\end{equation}
\begin{equation}
\delta P_{\; ,i}^{\alpha}=P_{\; ,ij}^{\alpha}Q_{\; \beta}^{j}
\delta \xi^{\beta}=\Bigr(C_{\; \beta \gamma}^{\alpha} P_{\; ,i}^{\gamma}
-P_{\; ,j}^{\alpha}Q_{\; \beta ,i}^{j}\Bigr)\delta \xi^{\beta},
\label{(3.4)}
\end{equation}
which imply 
\begin{equation}
\delta F_{ij}=-\Bigr(F_{kj}Q_{\; \alpha,i}^{k}
+F_{ik}Q_{\; \alpha,j}^{k}\Bigr)\delta \xi^{\alpha},
\label{(3.5)}
\end{equation}
and hence, bearing in mind Eq. (2.6),
\begin{equation}
\delta G^{ij}=\Bigr(Q_{\; \alpha,k}^{i}G^{kj}
+Q_{\; \alpha,k}^{j}G^{ik}\Bigr)\delta \xi^{\alpha},
\label{(3.6)}
\end{equation}
\begin{equation}
\delta {\widetilde G}^{ij}=\delta G^{+ij}-\delta G^{-ij}
=\Bigr(Q_{\; \alpha,k}^{i}{\widetilde G}^{kj}
+Q_{\; \alpha,k}^{j}{\widetilde G}^{ik} \Bigr)\delta \xi^{\alpha}.
\label{(3.7)}
\end{equation}
Equation (3.3), in particular, is obtained from the second Ward
identity \cite{DeWi03, DeWi84, DeWi65} 
for functional derivatives of the gauge-invariant action $S$:
\begin{equation}
S_{,ijk}Q_{\; \alpha}^{i}+S_{,ij}Q_{\; \alpha,k}^{i}
+S_{,ik}Q_{\; \alpha,j}^{i}+S_{,i}Q_{\; \alpha,jk}^{i}=0,
\label{(3.8)}
\end{equation}
jointly with the linearity of $Q_{\; \alpha}^{i}$ expressed by Eq. (1.5).
In the course of deriving the law (3.5), the four terms involving structure
constants from the use of gauge transformation laws (1.9) and (3.4) cancel
each other exactly.
    
By virtue of Eqs. (3.1), (3.2) and (3.7) we prove immediately gauge
invariance of the Peierls bracket (1.14), because
\begin{eqnarray}
\delta(A,B)&=& \Bigr[-A_{,k}Q_{\; \alpha,i}^{k}
{\widetilde G}^{ij}B_{,j}+A_{,k}Q_{\; \alpha,i}^{k}
{\widetilde G}^{ij}B_{,j} \nonumber \\
&+& A_{,i}Q_{\; \alpha,k}^{j}{\widetilde G}^{ik}B_{,j}
-A_{,i}{\widetilde G}^{ij}Q_{\; \alpha,j}^{k}B_{,k}\Bigr]
\delta \xi^{\alpha}=0. 
\label{(3.9)}
\end{eqnarray}

Since the sufficient condition (2.18) is very 
restrictive, it would be desirable
to use the sufficient condition (2.11) only while still being able to
prove gauge invariance of the Peierls bracket (1.14). This is indeed
possible because, in its final form (3.5), the gauge transformation law
for the gauge-field operator $F_{ij}$ is independent of the
functional derivatives $\omega_{\alpha \beta,i}$ and
$P_{\; ,ij}^{\alpha}$. We can therefore first assume that Eq. (3.5)
holds and later take $\omega_{\alpha \beta}$ to be a non-singular,
symmetric continuous matrix independent of field variables. 

\section{Towards a Moyal bracket on the space of histories}

Since the Peierls bracket (1.14) is a Poisson bracket, and bearing in
mind that the Poisson bracket of two functions on a manifold is the
coefficient of the term linear in $i {\hbar}$ in the corresponding
Moyal bracket \cite{Grac02}, we are now led to study how a Moyal bracket
among gauge-invariant functionals can be defined. Our starting point is
a careful consideration of a formula defining the star-product
of phase-space functions. Following the appendix of Ref. \cite{Grac02}, 
we recall that such a product may be expressed in the form
\begin{equation}
f \star g \equiv fg+{i{\hbar}\over 2}\left \{ f,g \right \}
+\sum_{k=2}^{\infty}\left({i {\hbar}\over 2}\right)^{k}
{1\over k!}D_{k}(f,g),
\label{(4.1)}
\end{equation}
where $D_{k}$ is a bidifferential operator defined by 
\begin{equation}
D_{k}(f,g)(q,p) \equiv \left[\left({\partial \over \partial q_{1}}
{\partial \over \partial p_{2}}
-{\partial \over \partial q_{2}}{\partial \over \partial p_{1}}
\right)^{k}f(q_{1},p_{1})g(q_{2},p_{2})
\right]_{q_{1}=q_{2}=q, p_{1}=p_{2}=p}.
\label{(4.2)}
\end{equation}
The star-product (4.1) may be recovered from the 
(asymptotic) expansion of
$$
f \; \exp \left[{i \hbar \over 2}\left(
{{\lvec \partial} \over \partial \xi^{j}} \omega^{jk}
{{\rvec \partial} \over \partial \xi^{k}}\right)\right] \; g,
$$
where $\xi^{i}$ takes the $2N$ values
$q^{1},...,q^{N},p_{1},...,p_{N}$. The Moyal bracket is eventually
obtained from the definition
\begin{equation}
[f,g]_{M} \equiv f \star g - g \star f
=i{\hbar} \left \{ f,g \right \}
+\sum_{k=2}^{\infty} \left({i \hbar \over 2}\right)^{k}
{1\over k!}\Bigr[D_{k}(f,g)-D_{k}(g,f)\Bigr],
\label{(4.3)}
\end{equation}
where even values of $k$ give vanishing contribution to the Moyal bracket.

In our field-theoretical framework, the inverse $\omega^{jl}$ of the
symplectic form (also denoted by $\Lambda^{jl}$) is replaced by the
supercommutator ${\widetilde G}^{jl}$, and we are led to the following
heuristic definition of the star-product of gauge-invariant functionals:
\begin{equation}
A \star B \equiv A \; \exp \left[{i \hbar \over 2} \left(
{{\lvec \delta} \over \delta \varphi^{j}}
{\widetilde G}^{jk} 
{{\rvec \delta} \over \delta \varphi^{k}}\right)\right] B.
\label{(4.4)}
\end{equation}
Upon expansion of the exponential, Eq. (4.4) yields, bearing in mind the
definition (1.14), 
\begin{eqnarray}
\; & \; & A \star B = AB+{i \hbar \over 2}(A,B)
+{(i \hbar /2)^{2} \over 2!} A_{,jl} {\widetilde G}^{jk}
{\widetilde G}^{lm} B_{,km} \nonumber \\
&+& {(i \hbar /2)^{2}\over 2!}\Bigr[A_{,jl}{\widetilde G}^{jk}
{\widetilde G}_{\; \; \; ,k}^{lm}B_{,m}
+A_{,j}{\widetilde G}_{\; \; \; ,l}^{jk}
{\widetilde G}_{\; \; \; ,k}^{lm} B_{,m} 
+ A_{,j}{\widetilde G}_{\; \; \; ,l}^{jk}
{\widetilde G}^{lm} B_{,km}\Bigr] \nonumber \\
&+& {\rm O}({\hbar}^{3}).
\label{(4.5)}
\end{eqnarray}
For our star-product to be associative, we have to {\it assume} that the
associative law of multiplication holds also for our contractions involving
both summation over repeated indices and integration over a space-time
region. Ultimately, this amounts to requiring a suitable rate of fall-off
at infinity or a suitable choice of boundary conditions (Section 4.1 of
Ref. \cite{DeWi84}).

Interestingly, the functional derivatives of 
${\widetilde G}^{lm}$ provide additional 
terms with respect to the formulae of ordinary quantum mechanics. However,
the gauge-invariant Moyal bracket 
\begin{equation}
[A,B]_{M} \equiv A \star B - B \star A 
\label{(4.6)}
\end{equation}
retains the same functional form as in ordinary quantum mechanics. Indeed,
on using the super-condensed DeWitt notation \cite{DeWi84} 
\begin{equation}
A_{,i} \equiv A_{1}, \; A_{,ij} \equiv A_{2}, ..., 
A_{,i_{1}...i_{l}} \equiv A_{l},
\label{(4.7)}
\end{equation}
one finds, from Eqs. (4.4)--(4.6),
\begin{eqnarray}
\; & \; & [A,B]_{M}=i{\hbar} (A,B)+{(i \hbar /2)^{2}\over 2!}\Bigr[
A_{2}{\widetilde G} {\widetilde G}B_{2}-B_{2} {\widetilde G}
{\widetilde G} A_{2} \nonumber \\
&+& A_{2} {\widetilde G} {\widetilde G}_{1}B_{1}
+A_{1}{\widetilde G}_{1}{\widetilde G}_{1}B_{1}
+A_{1}{\widetilde G}_{1}{\widetilde G}B_{2} \nonumber \\
&-& B_{2} {\widetilde G} {\widetilde G}_{1}A_{1}
-B_{1}{\widetilde G}_{1} {\widetilde G}_{1}A_{1}
-B_{1}{\widetilde G}_{1}{\widetilde G}A_{2}\Bigr]
+{\rm O}({\hbar}^{3}),
\label{(4.8)}
\end{eqnarray}
where exact cancellations occur among the various terms in square brackets 
in Eq. (4.8), because
\begin{eqnarray}
\; & \; & A_{2} {\widetilde G} {\widetilde G} B_{2}
-B_{2} {\widetilde G} {\widetilde G} A_{2} 
=A_{,jl}{\widetilde G}^{jk} {\widetilde G}^{lm} B_{,km}
-B_{,jl}{\widetilde G}^{jk}{\widetilde G}^{lm}A_{,km} \nonumber \\
&=& A_{,km}{\widetilde G}^{kj}{\widetilde G}^{ml}B_{,jl}
-B_{,jl}{\widetilde G}^{jk}{\widetilde G}^{lm}A_{,km}
=(-1)^{2}A_{,km}{\widetilde G}^{jk}{\widetilde G}^{lm}B_{,jl}
-B_{,jl}{\widetilde G}^{jk}{\widetilde G}^{lm}A_{,km} \nonumber \\
&=& B_{,jl} {\widetilde G}^{jk}{\widetilde G}^{lm}A_{,km}
-B_{,jl}{\widetilde G}^{jk}{\widetilde G}^{lm}A_{,km}=0,
\label{(4.9)}
\end{eqnarray}
\begin{eqnarray}
\; & \; & A_{2} {\widetilde G} {\widetilde G}_{1}B_{1}
-B_{1}{\widetilde G}_{1}{\widetilde G}A_{2}
=A_{,jl}{\widetilde G}^{jk}{\widetilde G}_{\; \; \; ,k}^{lm}B_{,m}
-B_{,j}{\widetilde G}_{\; \; \; ,l}^{jk}{\widetilde G}^{lm}A_{,km}
\nonumber \\
&=& A_{,km}{\widetilde G}^{kl}{\widetilde G}_{\; \; \; ,l}^{mj}B_{,j}
-B_{,j}{\widetilde G}_{\; \; \; ,l}^{jk}{\widetilde G}^{lm}A_{,km}
=A_{,km}{\widetilde G}^{kl}{\widetilde G}_{\; \; \; ,l}^{mj}B_{,j}
-B_{,j}{\widetilde G}_{\; \; \; ,l}^{kj} {\widetilde G}^{ml}
A_{,km}(-1)^{2}, \nonumber \\
&=& B_{,j}{\widetilde G}_{\; \; \; ,l}^{mj}{\widetilde G}^{kl}A_{,km}
-B_{,j}{\widetilde G}_{\; \; \; ,l}^{kj}{\widetilde G}^{ml}A_{,km}
=B_{,j}{\widetilde G}_{\; \; \; ,l}^{mj}{\widetilde G}^{kl}A_{,mk}
-B_{,j}{\widetilde G}_{\; \; \; ,l}^{kj}{\widetilde G}^{ml}A_{,km}
\nonumber \\
&=& B_{,j}{\widetilde G}_{\; \; \; ,l}^{kj}{\widetilde G}^{ml}A_{,km}
-B_{,j} {\widetilde G}_{\; \; \; ,l}^{kj} {\widetilde G}^{ml}A_{,km}=0,
\label{(4.10)}
\end{eqnarray}
\begin{eqnarray}
\; & \; & A_{1}{\widetilde G}_{1} {\widetilde G}B_{2}
-B_{2}{\widetilde G}{\widetilde G}_{1}A_{1}
=A_{,j}{\widetilde G}_{\; \; \; ,l}^{jk}{\widetilde G}^{lm}B_{,km}
-B_{,jl}{\widetilde G}^{jk}{\widetilde G}_{\; \; \; ,k}^{lm}A_{,m}
\nonumber \\
&=& A_{,m}{\widetilde G}_{\; \; \; ,k}^{ml}{\widetilde G}^{kj}
B_{,lj}-B_{,jl}{\widetilde G}^{jk}{\widetilde G}_{\; \; \; ,k}^{lm}
A_{,m}=(-1)^{2}B_{,jl}{\widetilde G}^{jk}
{\widetilde G}_{\; \; \; ,k}^{lm}A_{,m}
-B_{,jl}{\widetilde G}^{jk}{\widetilde G}_{\; \; \; ,k}^{lm}A_{,m}
\nonumber \\
&=& 0,
\label{(4.11)}
\end{eqnarray}
\begin{eqnarray}
\; & \; & A_{1}{\widetilde G}_{1} {\widetilde G}_{1}B_{1}
-B_{1}{\widetilde G}_{1} {\widetilde G}_{1}A_{1}
=A_{,j}{\widetilde G}_{\; \; \; ,l}^{jk} {\widetilde G}_{\; \; \; ,k}^{lm}
B_{,m}-B_{,j}{\widetilde G}_{\; \; \; ,l}^{jk}
{\widetilde G}_{\; \; \; ,k}^{lm}A_{,m} \nonumber \\
&=& A_{,m}{\widetilde G}_{\; \; \; ,k}^{ml} 
{\widetilde G}_{\; \; \; ,l}^{kj} B_{,j}
-B_{,j}{\widetilde G}_{\; \; \; ,l}^{jk}
{\widetilde G}_{\; \; \; ,k}^{lm}A_{,m}
=(-1)^{2}A_{,m}{\widetilde G}_{\; \; \; ,k}^{lm}
{\widetilde G}_{\; \; \; ,l}^{jk}B_{,j}
-B_{,j}{\widetilde G}_{\; \; \; ,l}^{jk}
{\widetilde G}_{\; \; \; ,k}^{lm}A_{,m} \nonumber \\
&=& B_{,j}{\widetilde G}_{\; \; \; ,l}^{jk}
{\widetilde G}_{\; \; \; ,k}^{lm}A_{,m} 
-B_{,j}{\widetilde G}_{\; \; \; ,l}^{jk}
{\widetilde G}_{\; \; \; ,k}^{lm}A_{,m}=0.
\label{(4.12)}
\end{eqnarray}
In the course of deriving Eqs. (4.9)--(4.12), besides relabelling
dummy indices and exploiting anti-symmetry of ${\widetilde G}^{jk}$
and commutation of second functional derivatives for type-I theories,
we have {\it assumed} that the associative law of multiplication holds
in agreement with the assumption made following Eq. (4.5).

We have therefore proved explicitly that the definition (4.6) of 
gauge-invariant Moyal bracket may engender the asymptotic expansion
\begin{equation}
[A,B]_{M}=i{\hbar}(A,B)+{\rm O}({\hbar}^{3}).
\label{(4.13)}
\end{equation}
It remains to be seen whether, order by order in $\hbar$, the functional
derivatives of the supercommutator function ${\widetilde G}^{jk}$ give
always vanishing contribution to $[A,B]_{M}$. This can be done by hand
with finitely many terms in the expansion of Eq. (4.4), but more powerful 
methods are necessary to obtain a proof to all orders.

\section{Concluding remarks and open problems}

We have studied structural issues which are relevant for the manifestly
covariant approach to quantization of gauge theories. In Secs. II and III
we have put on firm ground the choice of linear covariant gauges from the
point of view of Peierls-bracket formalism: the Jacobi identity and the
choice of generic gauge fields on the space of histories enforce the 
choice of linear covariant gauges. In Sec. IV we have defined, by
close inspection of the quantum mechanical formalism, 
a Moyal bracket on the space of histories (which is new to our
knowledge), proving that it reduces to $i {\hbar}$ times 
the Peierls bracket to lowest order
in $\hbar$. Several important problems deserve now investigation, i.e.
\vskip 0.3cm
\noindent
(i) How to replace the definition (4.4) by exploiting a suitable
integral kernel, along the lines of Appendix A of Ref. \cite{Grac02},
in such a way that Moyal brackets for type-I gauge theories are put on
firm ground, without relying upon formal series.
\vskip 0.3cm
\noindent
(ii) How to apply such a formalism to quantized general relativity
\cite{DeWi60}.
\vskip 0.3cm
\noindent
(iii) How to define Peierls brackets for noncommutative extensions of
general relativity, bearing in mind the work in Refs. 
\cite{Asch05,Asch06}.
\vskip 0.3cm
\noindent
(iv) What is the relation, if any, with modern non-perturbative 
approaches to quantum gravity \cite{Rove04}.

Hopefully, the years to come will tell us whether a renaissance of
Peierls-bracket formalism may lead to a better understanding of the
difficulties faced by any attempt of quantizing the gravitational field
\cite{DeWi03}. Note also that, if the Moyal bracket is viewed as being
more fundamental, Eq. (4.13) suggests {\it defining the Peierls bracket}
from the relation
\begin{equation}
(A,B) \equiv \lim_{\hbar \to 0}{1\over i \hbar}[A,B]_{M},
\label{(5.1)}
\end{equation}
which provides, to our knowledge, a novel way of looking at the
Peierls bracket.

\acknowledgments
Previous collaboration with Giuseppe Bimonte and Giuseppe Marmo on
related topics has taught us a lot and has provided the appropriate
motivation for the present research. The authors are also grateful to 
Fedele Lizzi and Patrizia Vitale for clarifying the work in 
Ref. \cite{Grac02}, and to the INFN for
financial support. Their work has been partially
supported by PRIN {\it SINTESI}.

\end{document}